%% file: papermn1.tex
\documentclass[letters,usenatbib,fleqn]{mnras}
\usepackage{newtxtext,newtxmath}
\usepackage[T1]{fontenc}
\usepackage{ae,aecompl}

\usepackage{graphicx}	
\usepackage{amsmath}	
\usepackage{amssymb}	
\usepackage{bm,url,times}
\usepackage{float}
\def\Pm{\mbox{\rm P}_M}
\def\Rm{\mbox{\rm R}_M}

\def\Brmso{B_{\rm rms0}}
\def\Urmso{U_{\rm rms0}}
\def\apj{ApJ}

\def\mnras{MNRAS}
\def\aap{A\&A}
\def\apjl{ApJ}

\def\pre{PRE}

\def\apjs{ApJS}

\usepackage{bm}
\usepackage{times}
\usepackage{dcolumn}
\usepackage{epsfig}

\input macros.tex
\graphicspath{{./}{./png/}}

\title[]{Evolution of the magnetorotational instability on initially tangled magnetic fields}
 
\author[Bhat et. al.]{Pallavi Bhat,$^{1,2}$\thanks{E-mail: pbhat@mit.edu} 
Fatima Ebrahimi,$^{1}$
Eric G. Blackman,$^{3}$
and Kandaswamy Subramanian,$^{4}$
\\
$^{1}$Department of Astrophysical Sciences and Princeton Plasma Physics Laboratory, Princeton University, Princeton, NJ 08543, USA\\
$^{2}$Plasma Science and Fusion Center, Massachussetts Institute of Technology, Cambridge, MA 02139, USA\\
$^{3}$Department of Physics and Astronomy, University of Rochester, Rochester, NY 14618, USA\\
$^{4}$Inter University Centre for Astronomy and Astrophysics,Post Bag 4, Pune University Campus, Ganeshkhind, Pune 411 007, India
}

\date{Accepted XXX. Received YYY; in original form ZZZ}

\pubyear{2017}

\begin{document}
\label{firstpage}
\pagerange{\pageref{firstpage}--\pageref{lastpage}}
\maketitle

\begin{abstract}
The initial magnetic field  of previous magnetorotational instability (MRI) simulations 
 has always included  a significant system-scale component, even if stochastic.  
However, it is of  conceptual and  practical interest to assess whether 
 the MRI can grow when the initial field is turbulent.
The ubiquitous presence of turbulent or random flows in astrophysical plasmas  
generically leads to a small-scale dynamo (SSD), which would provide  initial seed  turbulent velocity and magnetic fields in the plasma that becomes an accretion disc.
Can the MRI  grow  from these more realistic   initial conditions?
To address this  we supply a standard shearing box with isotropically forced
SSD generated magnetic and velocity fields as initial conditions, and remove the 
forcing.
We find that if the initially supplied fields are too weak or too incoherent,
they decay from the initial turbulent cascade faster than they can grow
via the MRI.  When the initially supplied fields are sufficient to allow MRI growth and sustenance,  
the saturated stresses, large-scale fields, and power spectra 
match those of the standard zero net flux MRI simulation with an initial  
large scale vertical field.
\end{abstract}



\section{Introduction}

Magnetic fields are ubiquitous in turbulent astrophysical plasmas.
While the mechanism of origin of large-scale ordered magnetic fields in these
systems is a subtle business, more generic and less controversial is the  
amplification of total magnetic energy by the fluctuation or small-scale dynamo (SSD). 
Here, turbulence in a conducting plasma, with even a modest magnetic Reynolds number ($\Rm$), 
leads to exponential growth 
of the field
on the shortest eddy turn over time scales, 
which is usually much smaller than the age of the
astrophysical system.
The SSD is likely to be important 
for the early generation of magnetic fields in stars and galaxies/inter-stellar medium (ISM) \citep{Suretal10,gentetal13,BS13}.
Such SSD generated fields would then be present in
the plasma from stars or the ISM that source accretion disks.

In previous studies of the magneto-rotational instability (MRI) in shearing box models of accretion disks,  
the initial condition is typically an ordered \textit{non-stochastic}  field  with net flux or zero-net flux. 
Following linear stability analysis, this kind of initial condition is the most natural to compare with 
minimalist analytic theory, but in reality one expects a more random field
without necessarily much of a large-scale field.
The question then arises as to whether and what kind of disorder  
and incoherence in the initial field can still lead to MRI and field growth. 
Moreover, in general the source of an initial magnetic field itself, which could trigger MRI in a disk is not known. 
Initial fields in disks may be supplied externally and this may also be generated by SSD.
Here, we explore this latter possibility, and the condition under which 
large-scale fields could grow and be sustained through MRI-driven turbulence initiated by SSD fields.

\begin{figure*}
\includegraphics[width=0.495\textwidth, height=0.3\textheight]{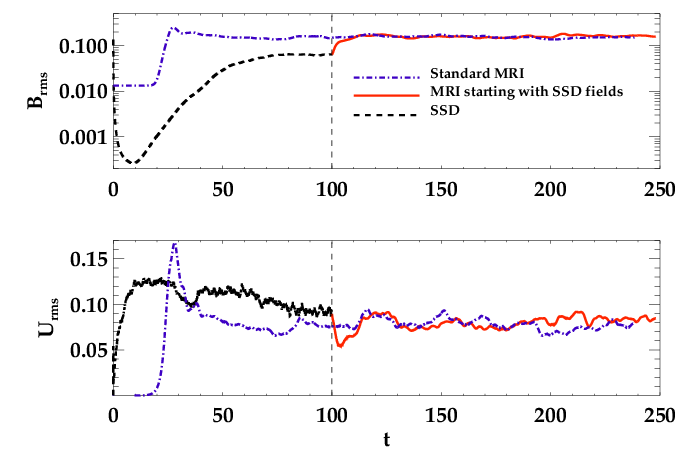}
\includegraphics[width=0.495\textwidth, height=0.3\textheight]{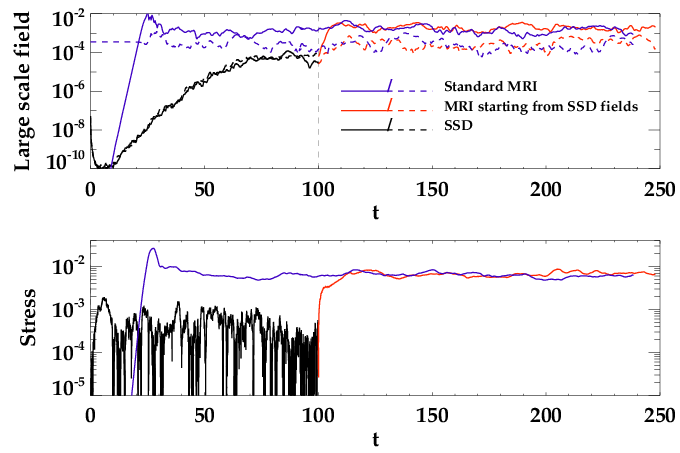}
\caption{Evolution of $\Brms$ and $\Urms$ is shown in upper and lower left panels respectively for Run~SSD, Run~A0 and Run~STD.
Evolution of energy in large-scale magnetic fields and sum of Maxwell and Reynold stresses 
is shown in upper and lower right panels respectively for Run~SSD, Run~A0 and Run~STD. 
The solid curves refer to $x$-$y$ averaged fields while the dashed curves refer to $y$-$z$ averaged fields.
}
\label{timeplot}
\end{figure*}

In the case of initial vertical fields of zero flux or net flux field, the MRI modes grow large-scale radial and azimuthal fields 
in the early linear growth phase \citep{BEB16, EB16} on orbital times scales 
eventually saturating nonlinearly to generate turbulence. It has been a long-standing topic of 
investigation to understand what determines the amplitude of the fields and stresses on MRI saturation, because this
is thought to constrain the rate of angular momentum transport in accretion disks. 
The saturation amplitude of the stresses is not a constant across simulations but, for fixed resolution and box dimensions, depends on whether 
the initial field is of zero-net flux type or there is a uniform background field \citep{HGB95, fromang07, Guanetal2009, Shi2016}. 
 
Here we investigate whether random fields (without a significant large-scale component) can seed the MRI, and whether the MRI sustains.
Previous work using initially random fields \citep{HGB96} had adopted a flat
1-D magnetic spectrum, and also with mostly large scale modes having
$1 < k(L/2\pi) < 4$ and the field being $0$ outside that interval. Such an initial field has a significant large
scale component, (with the box scale field comparable to smaller 
scale fields), unlike the initial conditions we adopt below.
In particular, we show that using small-scale fields 
(which by definition implies the absence of significant power at large scales)
from a small-scale dynamo (SSD) as an initial condition, 
subsequent shear and rotation act to trigger MRI and as a result, sustains the turbulent fields even 
after the SSD forcing is turned off.
We also discuss that the MRI fails to sustain when Gaussian random noise is instead used as an initial field.
Note that \citet{Riols13}
generate an initial condition with random set of Fourier modes to
study the transition to MRI turbulence,
but with apparently large scale modes being energetically dominant. In our case, the Gaussian random initial fields are essentially noise-like 
with no large-scale component.

In section 2, we compare the sustained MRI turbulence with SSD initial condition 
against the case with zero net flux initial condition, with respect to  the standard MRI signatures. 
Section 3 discusses the role of coherence in initial random magnetic fields for MRI to grow. The 
nature of MRI starting with SSD generated
fields is investigated by spectral analysis in Section 4. We
conclude  in Section 5.

\section{Sustained MRI turbulence with SSD Initial Condition}
We perform shearing box simulations in a periodic box of $1:1:1$ aspect ratio using the \textsc{Pencil Code}\footnote{https://github.com/pencil-code \citep{B03}}. 
Most MRI simulations are of the resolution of $256^3$ grid points and one fiducial run with $512^3$ grid points. 
The model and the details of the simulation setup are the same as in 
\citet{BEB16}.
The simulations are performed in two stages. First, 
to obtain the initial condition, we run a simulation of the SSD, without
shear or rotation, with an imposed isotropic stochastic forcing in the momentum equation. Second, 
the velocity and magnetic fields 
from this small-scale dynamo simulation (either from the kinematic 
or staturated stage), 
are introduced into a shearing box simulation as initial condition, 
with uniform shear $U_y=Sx$ (where $S$ is the shearing rate) and rotation. The rotation is obtained by adding the 
Coriolis force term in the momentum equation, $2({\bf \Omega}\times\UU)$, where ${\bf \Omega}=\Omega_0 \zzz$.
In all of our runs, $S=-1.5$ and $\Omega_0=1$. 
Note that the stochastic forcing term used to obtain the SSD saturated state is subsequently turned off 
when the shearing box simulation begins with the SSD generated field
as its initial condition.
In principle, we could have run our SSD simulation in presence of shear, but then the properties 
of the SSD fields to be used as initial condition for MRI runs would vary as a function of the shearing rate. 
Thus we keep it simpler and the only free parameter which changes the properties of the SSD fields is the forcing scale. 

In  \Fig{timeplot}, we show the evolution of root mean square  magnetic and velocity fields over time, 
for three runs of $512^3$ resolution,  
(i) The SSD simulation 
run to saturation
which is used to set the initial condition, called Run~SSD
(ii) the fiducial shearing box MRI simulation with initial fields from 
this SSD simulation, referred to as Run~A0, starting at $t\sim120$; 
and (iii) a standard  MRI run starting with zero net flux of vertical large scale mode, Run~STD.   
The magnetic Reynolds number is defined as $\Rm=\Urms L/\eta$, where $L$ is the size of the box and $\eta$
is the microscopic resistivity and $\Rm=12000$ for Run~SSD, while in Run~STD and Run~A0, the resulting $\Rm$ is $6000$.
The Prandtl number is $\Pm=\nu/\eta$ and $\Pm=10$ for all the three runs.  
It can be seen that for Run~A0 (shown in solid red lines), initial saturated SSD fields do not
decay but grow and sustain due to MRI. 
The $\Brms$ grows by a factor of $\sim 2$ to a value $\sim 0.13$ 
and $\Urms$ decays by a factor of $\sim2$,  saturating at a value of $\sim 0.06$. 
Interestingly the saturation levels of both $\Brms$ and $\Urms$ match with those from Run~STD.
\Fig{timeplot} also shows that the $\Urms$ decays 
to a value which is smaller than the $\Brms$, thus going 
from a kinetically dominated system (SSD saturation phase) to a magnetically dominated system as would be the case for MRI turbulence.

We show the  Reynolds and Maxwell stresses 
$\bra{U_x U_y}$ and $\bra{B_x B_y}$  for  Run~A0 compared to Run~STD in the 
lower right panel of 
\Fig{timeplot}.
The amplitude of the sum of the stresses in Run~A0 is $\sim 0.005$, which is the same 
as in Run~STD.  
On the other hand, the stresses in Run~SSD
are very noisy and  orders of magnitude below those of  Run~A0. 
A distinctive increase at $t\sim120$ is seen, indicating the presence of MRI instability.
We find that the Maxwell to Reynolds stress ratio in Run~A0 is about $6.9$, which matches well with
the ratio $\sim 6.7$ estimated from the standard MRI simulation, Run~STD \citep{Axel95}. 

In the upper right panel of \Fig{timeplot}, 
we also show the evolution of planar averaged large-scale fields.
The solid lines refer to $x$-$y$ averaged fields and the dashed lines show the $y$-$z$ averaged fields.
The growth in the large-scale field in Run~SSD is due to the low wavenumber tail 
of the magnetic power (peaked at large $k$)  that results because  the SSD eigenfunction grows self-similarly  
across all scales \citep{BS13, SB14, BSB16}. However the saturation level is rather small at $~10^{-5}$.
In the standard MRI simulation, the large-scale dynamo amplifies low wavenumber fields to a much higher amplitude 
$\sim 10^{-3}$ \citep{BEB16},
and this also obtains in Run~A0.
In fact, from \Fig{timeplot} top right panel, we 
see that again the amplitude of energy in large-scale fields from Run~A0 matches with Run~STD.

\section{Importance of Coherence Scales}

\begin{table}
\begin{center}
\setlength{\tabcolsep}{3.5pt}
\begin{tabular}{|l|c|c|c|c|c|c|}
\hline
\hline
\textbf{Run} & $\kf$ & $k_{int}$ & $\Brmso$ & $\Urmso$ & $\gamma_{\Brms}$ & Resolution\\
\hline
\hline
 &  &  & &  &  & \\
A0& 1.5 & 7 & 0.06  & 0.090 & 0.0320 & $512^3$\\	
A & 1.5 & 7 & 0.06  & 0.095 & 0.0274 & $256^3$\\	
B & 5 	& 14  & 0.06  & 0.09  & 0.0088 & $256^3$\\	
C & 10 	& 23  & 0.05  & 0.08  & 0.0053 & $256^3$\\	
D & 1.5 & 14 & 0.02  & 0.12  & 0.0158 & $256^3$\\	
E & 1.5 & 15 & 0.008 & 0.12  & 0.0153 & $256^3$\\	
F & 1.5 & 15 & 0.004 & 0.12  & 0.0146 & $256^3$\\	
G & 1.5 & 15 & 0.002 & 0.12  & - 	 & $256^3$\\	
H & 25  & 40  & 0.03 & 0.07  & - 	 & $256^3$\\	
\hline
\hline
\end{tabular}
\end{center}
\caption{Summary of all runs: For each Run, the forcing scale $\kf$ of the SSD run from which 
the initial condition is taken is specified, along with the $k_{int}$, the magnetic integral scale. 
Also shown are the initial magnetic and velocity field strengths,
$\Brmso$ and $\Urmso$, the growth rate of $\Brms$ given by $\gamma_{\Brms}$ and the resolution.}
\label{Tab}
\end{table}

\begin{figure*}
\centering
\includegraphics[width=0.495\textwidth, height=0.27\textheight]{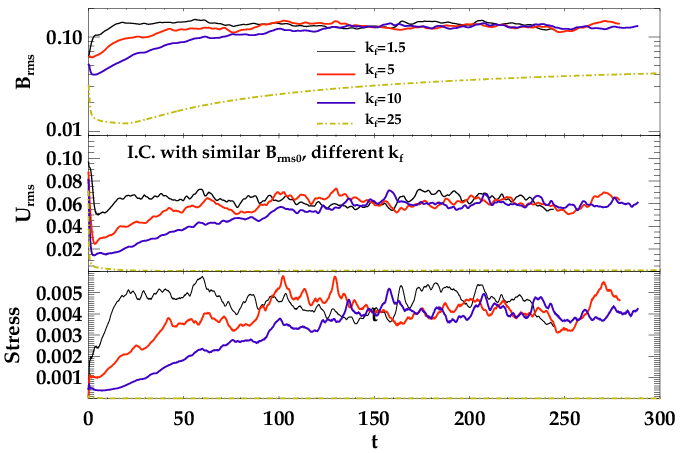}
\includegraphics[width=0.495\textwidth, height=0.27\textheight]{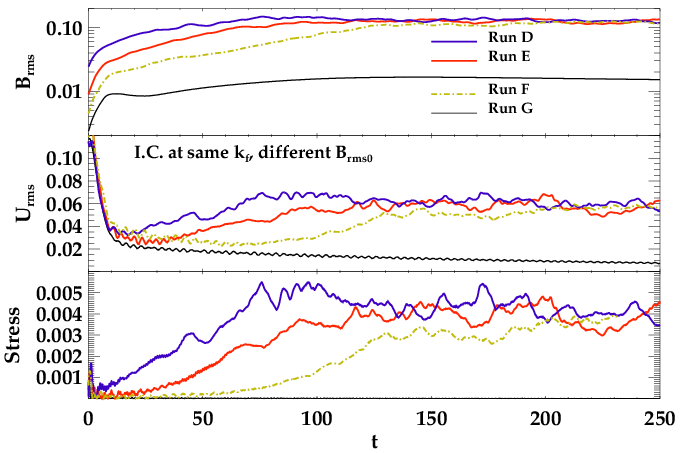}
\caption{Evolution of $\Brms$ and $\Urms$ and stresses is shown in upper, middle and lower panels respectively 
for varying $\kf$ in initial SSD fields in the left set and for varying $\Brmso$ in the right set.
In the left set: the black, red, blue and green curves correspond to Run~A, Run~B, Run~C and Run~H where initial fields have 
$\kf=1.5$, 5, 10 and 25 respectively with similar $\Brmso$ around $\sim 0.03--0.06$.
In the right set: the blue, red, dashed green and black curves correspond to Run~D, Run~E, Run~F and Run~G 
where initial fields have $\kf=1.5$, but decreasing $\Brmso$
We do not show the initial SSD phase here.
}
\label{diffkf}
\end{figure*}
\subsection{SSD vs. Gaussian random noise as Initial Conditions}
Unlike the case of the SSD initial conditions discussed in the previous section, 
we find  that  for a Gaussian random noise of even a large 
initial RMS field strength $\Brms \sim 0.1$ (finite perturbation), 
the field and velocity fluctuations decay 
in the shearing box run. This Gaussian random noise seed field was obtained by setting the 
vector potential to be normally distributed, uncorrelated random numbers in all meshpoints for all three components, 
which results in a spherical shell averaged magnetic power spectrum, $\EEM$, that increases with wave number $k$ as $\EEM\sim k^4$.
Note on the other hand, such Gaussian random noise field used as the initial condition 
in the Run~SSD, does indeed lead to field growth. 
This is a particularly important point to note, because 
it immediately shows that while SSD mechanism is robust to any type of initial condition, 
MRI is sensitive to the nature of the initial field. 
Also we find that even when the criterion of providing finite (large amplitude) perturbations is satisfied 
as required for subcritical transition to turbulence, MRI modes still do not grow. We find another criterion
that plays an important role namely the coherence scale of the initial fields. 
To understand this better, we examine the effect of varying magnetic integral scale (or the typical coherence scale) of the initial fields on MRI. 
The magnetic integral length is defined as, 
$L_{int}=(\int (2\pi/k)\EEM dk)/(\int \EEM dk)$. The corresponding wavenumber is given by $k_{int}=2\pi/L_{int}$

In the SSD simulations, the forcing term in the momentum equation drives vortical motions localised around a
wavenumber $\kf$, changing direction and phase at every time step (more details are given in \citet{HBD04}).
Thus the turbulent outer scale of the velocity field is set at $k=\kf$.
The growing magnetic fields in the SSD, peak near the resistive scales in the kinematic phase; 
but by saturation, the power shifts to larger scales closer to $\kf$ \citep{BS13}. 
Therefore, we investigate the effect of increasing $\kf$ (which increases $k_{int}$) during the SSD phase
and how this influences subsequent growth and sustenance of MRI turbulence once the stochastic forcing
is turned off.

\subsection{Sensitivity of MRI to the $\kf$ or $k_{int}$ of the initial  saturated SSD fields}
\begin{figure*}
\centering
\includegraphics[width=0.96\textwidth, height=0.25\textheight]{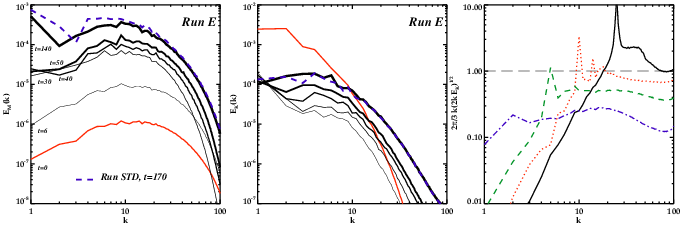}
\caption{Evolution of magnetic power spectrum is shown in the left panel at times $t=6,30,40,50$ and $140$ in
solid black lines of increasing thickness. Similarly kinetic power spectrum is shown in the middle panel
at times $t=30,40,50,60$ and $140$ in solid black lines of increasing thickness. These are shown for Run~E.
The red curves are the initial spectra. And the dashed green curves correspond to the spectra from Run~STD in saturation.
In the right panel, the curves for rate of turbulent diffusion for the initial velocity field, $kv(k)=k\sqrt{2k E_{K}(k)}$
are shown for Run~A (blue dash-dotted), Run~B (green dashed), 
Run~C (dotted) and Run~H (black). The dashed horizontal line is MRI growth timescale $\Omega=1$. 
Only Run~H doesn't sustain as the turbulent diffusion rate is larger than the MRI growth rate. 
}
\label{spec}
\end{figure*}

 \Fig{diffkf}  shows the time evolution of $\Brms$, $\Urms$ and stresses in MRI
simulations with initial SSD fields whose forcing wavenumbers were $\kf=1.5$, $5$, $10$ and $25$, 
indicated by black, blue, red and green lines respectively. 
Runs with $\kf=1.5$, $5$, $10$ and $25$ are referred to as Run~A, Run~B, Run~C and Run~H respectively.
The initial fields in the Runs~A, B , C, H have the respective magnetic integral wavenumbers, $k_{int}=7$, $14$, $23$ and $40$. 
The plots show that increasing $\kf$ decreases the average growth rate and thus increases the time to reach saturation.   
The MRI modes which are triggered must be of larger wavenumber $k$ for initial 
fields of larger $\kf$. 
One can perhaps understand this as follows:
For an initially uniform vertical field $B_0$, MRI unstable modes 
have a maximum growth rate for a wavenumber $k=k_{max} \propto \Omega/B_0$.
One may roughly adopt a similar estimate for random fields, with now
$B_0$ a measure of a suitably defined local coherent component of the field 
(such $B_0$ can be thought of as an average over a fixed scale between the different runs).
For larger $\kf$, the magnetic power from the SSD is also 
peaked at a larger $k$, and for the same RMS value of the field, 
the $B_0$ would be smaller or $k_{max}$ would be larger.
These larger MRI unstable $k$ modes are more affected by turbulent diffusion
than lower $k$ modes
and thus incur smaller growth rates. 
In essence, we require that the MRI growth rate $\Omega$ for a given mode be larger than 
the corresponding turbulent diffusion rate, $(2\pi/3)k v(k)$, where $v(k)$ is the velocity at the wavenumber $k$, 
obtained from the kinetic power spectrum as $\sqrt{2k E_{K}(k)}$ 
i.e. $\Omega > (2\pi/3) k v(k)$, where $k$ correspond to 
the fastest growing modes. We show in the right panel of \Fig{spec}, the curves $(2\pi/3) k v(k)$ from the initial velocity fields, 
for Runs~A, B, C and H. We only make the assumption that the expected $\kmax$ is either equal to or larger than $\kf$. 
Only in the case of Run H, the growth of MRI modes is ineffective as the horizontal line of $\Omega=1$ 
intersects the rate of turbulent diffusion already at about $k\sim20$.
The above interpretations also explain why an initial seed field of Gaussian random noise doesn't trigger MRI; namely that 
the coherence scale would be close to the grid resolution scale. 
Therefore, the presence of sufficiently coherent structures (effected here by small enough $\kf$) in SSD fields enables MRI growth,
thus implying a minimum field coherence requirement.
 
\begin{figure*}
\centering
\includegraphics[width=0.3\textwidth, height=0.2\textheight]{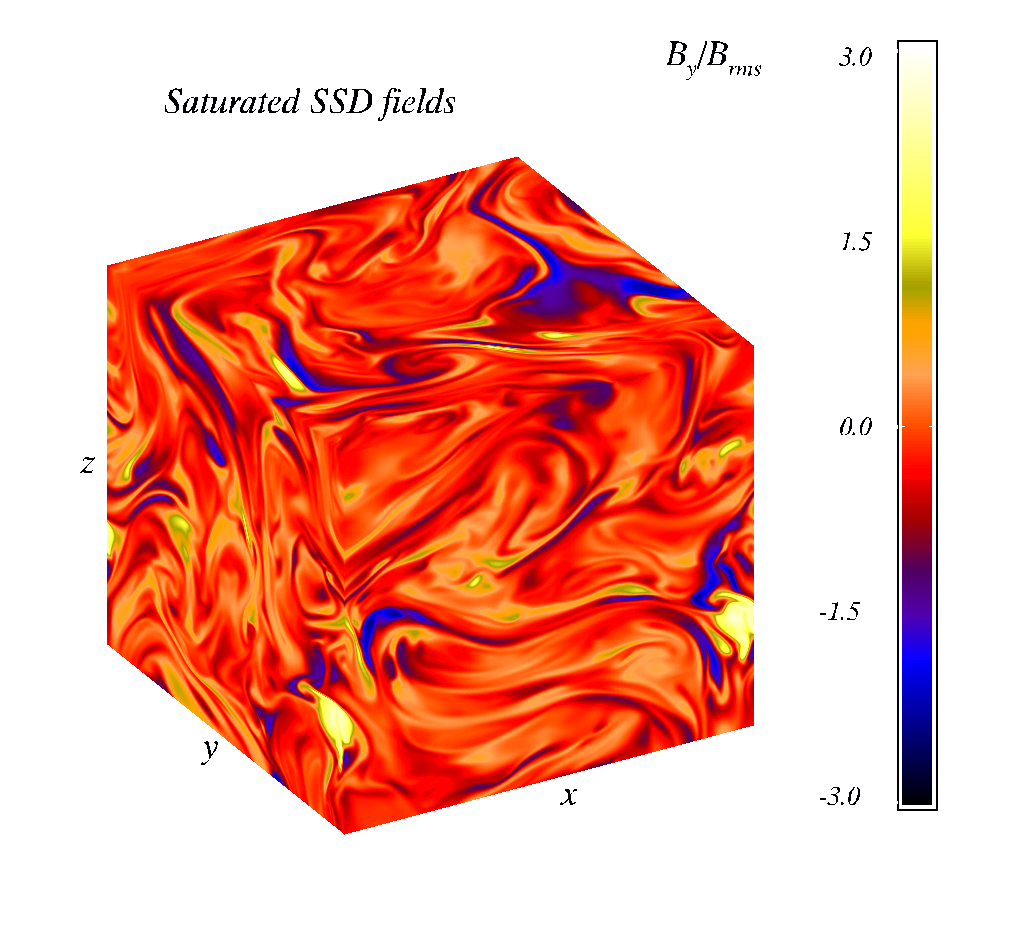}
\includegraphics[width=0.3\textwidth, height=0.2\textheight]{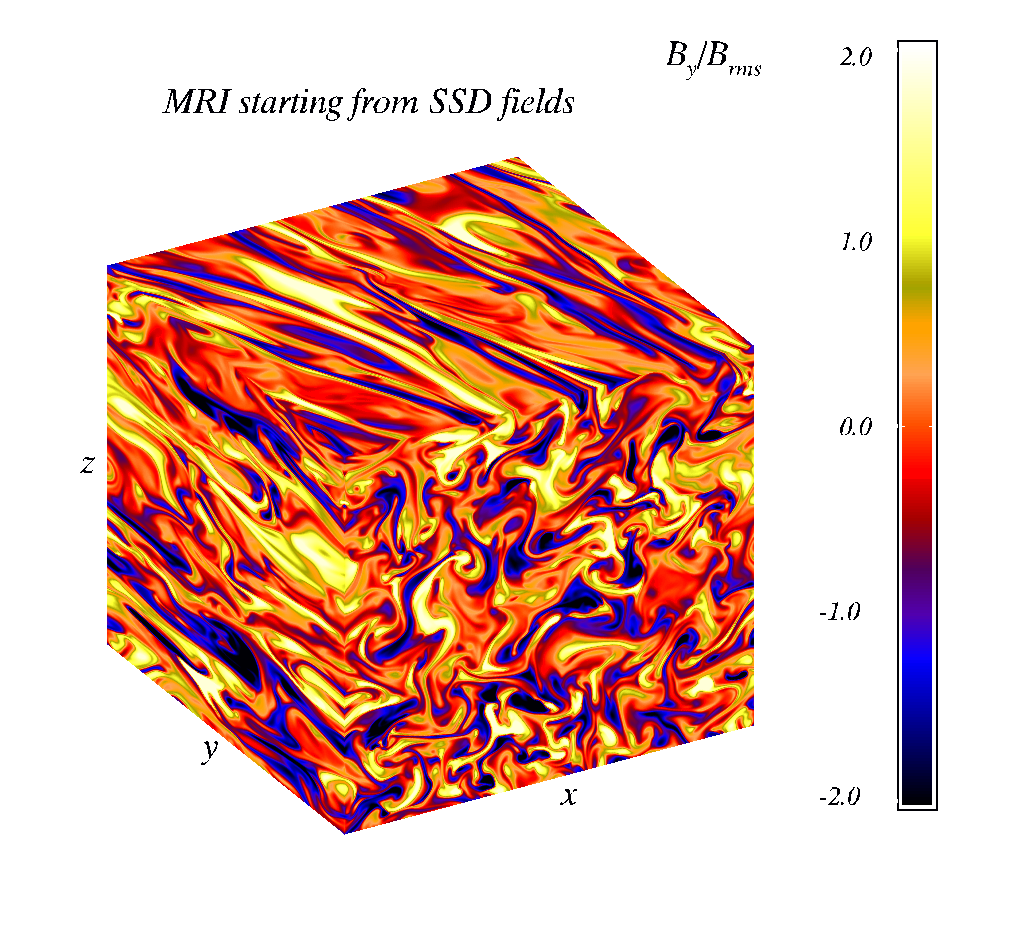}
\includegraphics[width=0.3\textwidth, height=0.2\textheight]{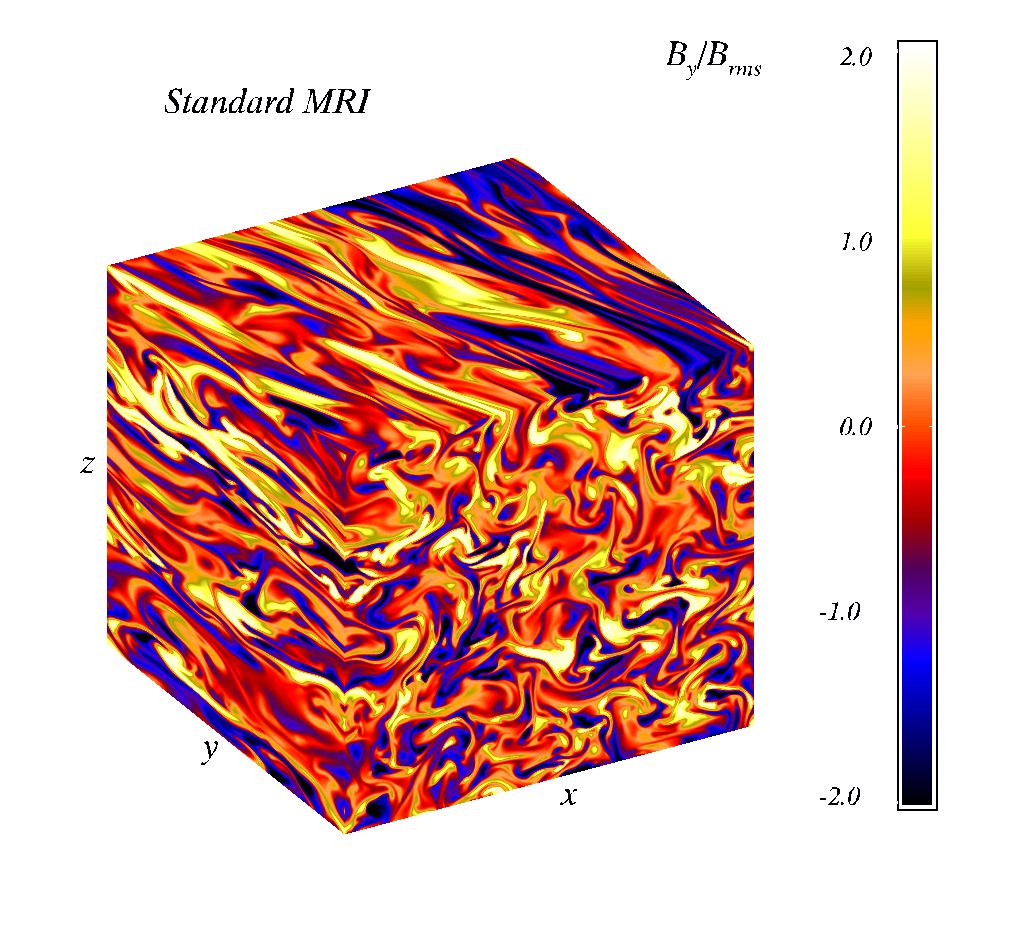}
\caption{The azimuthal or $B_y$ component of the magnetic fields are plotted for Run~SSD, Run~A0 and 
Run~STD at high resolution of $512^3$ in left, middle and right panels, from the saturated regimes. 
}
\label{field}
\end{figure*}

%
\subsection{Sensitivity of MRI to the strength of initial kinematic SSD fields}
While the saturated SSD fields seem to be sufficiently coherent to trigger the MRI
for a certain range of $\kf$  
we can also ask whether SSD fields from just the kinematic regime
are locally coherent enough to trigger MRI? 
We investigate this here using fields from the kinematic regime of SSD as initial condition for MRI.

The right panel of \Fig{diffkf} shows the time evolution of total 
$\Brms$, $\Urms$ and stresses in MRI 
simulations with SSD generated fields from the kinematic regime,
of different initial field strengths $\Brmso$, but the same $\kf=1.5$. 
The different curves are for  $\Brmso\sim 0.02$, 0.008, 0.004, 0.002, shown in 
solid blue, solid red, dash-dotted green and solid black respectively; also referred as Run~D, Run~E, Run~F and Run ~G.
We find that MRI is triggered for initial SSD fields even from its kinematic regime. 
Again, here we find a trend of decreasing growth rate with decreasing $\Brmso$. 
Note that as we decrease the initial field,  
we also decrease the effective $B_0$, and so 
increase the wavenumber of the fastest growing mode, where damping effects can slow down the growth. 
 
Note that the magnetic field growth happens in two regimes, as seen from the evolution curves 
of $\Brms$ and $\Urms$
in the right panels of \Fig{diffkf} .  
At first, even as $\Urms$ decays due to rapid turbulent cascade of the velocity,
$\Brms$ actually grows rapidly  due to enhanced random shearing from these velocity flows.
By "enhanced random shearing", we mean that even though the velocity field is decaying, the flow is still turbulent
and there is a temporary SSD action enhanced by the uniform shear \citep{SRB16}. But this gets subsequently overtaken by MRI action. 
Thereafter both $\Brms$ and $\Urms$ grow together 
due to MRI at the same rate up unto saturation. 
In the bottom-most curve in upper right panel of \Fig{diffkf} in black, there is an initial increase of the magnetic field 
due to the random shearing, which however is not enough to trigger modes which are sufficiently large 
scale to compete with diffusion. 
Thus MRI modes do not grow in this particular run with $\Brmso \sim 0.002$. And while the velocity field ($\Urms$) decays further, 
the magnetic field, after the initial rapid growth due to random shearing,
turns to saturate but continues to grow at a smaller rate due to the linear shearing action. However it eventually decays as it gets stretched out to resistive scales. 
We therefore show that the strength of total initial turbulent fields also has a critial role in triggering the MRI here.

Additionally, we considered the case in which SSD fields were allowed to decay (by removing the external forcing). 
Then such decayed (not fully) fields were used as initial condition for MRI. We again found that the MRI grows and 
saturates to the same amplitude as the other MRI growth cases. This is conceptually 
motivated by  the 
possible circumstance whereby a disk forms from the collapse of 
a turbulent cloud or object whose forcing may not survive the transition.
We have summarised the initial condition parameters for the different runs studied, in the \Tab{Tab}. 
The growth rate of the $\Brms$ calculated for each run is an average quantity given by $\gamma_{\Brms}=(\int_0^T dt~ d(\ln{\Brms})/dt)/T$,
where $T$ is the time taken to reach saturation. 

\section{Spectral analysis of MRI growth}
To understand how
the power spectrum of the turbulent initial field evolves once the MRI takes over
 and whether there are any modal signatures akin
to standard MRI linear phase, we study the initial SSD fields to MRI spectral evolution.
\Fig{spec} shows the evolution of magnetic and kinetic power spectra in Run~E 
(corresponding to the solid red curves in right panel of \Fig{diffkf})  in the left and right panels respectively.
The magnetic power spectrum, $\EEM$, first grows self-similarly from $t=0$ to $t=5$, due to 
enhanced random shearing action (similar to how the SSD spectrum grows  c.f. \cite{schek}, \cite{HBD04}). 
Then during 
$t=30$ to $t=50$, the growth is due to linear MRI modes (when both $\Brms$ and $\Urms$ grow together). 
While growth is expected at all $k$, 
the MRI modes $k\sim1$--$3$ do not grow much. Even if they did correspond to maximally growing modes, they would transfer all of their energy to larger wavenumbers. 
This indicates why the larger wavenumbers all grow by the same amount.  
A similar picture is seen for the kinetic spectra, $\EEK$, with little growth at lower wavenumbers ($k\sim1$--$3$),
but higher wavenumbers growing in unison. 
Note that between $t=50$ to $140$, the small $k$ modes become more and more prominent. 
In particular, the growth and maximum magnetic energy of small-$k$ modes is around the same time (around $t=140$) that the MRI stresses start to saturate (shown in Fig. 2 for run E).
This is consistent with the notion that MRI saturation is connected to large-scale 
magnetic field generation \citep{Ebrahimi2009}.  
We therefore find that starting from  more realistic turbulent SSD fields also sheds 
lights on the MRI saturation mechanism. Further analysis of MRI satutation remains for future work.
Finally, the thickest solid black curves in both panels 
from the saturated regime are compared with the 
dashed curve at a similar time from the standard MRI run and they match well. 

Note the peak at $k=1$, which also indicates presence of large-scale fields (besides planar averaging). 
We find that the peak at $k=1$ appears and disappears periodically, 
consistent with  temporal cycles in the large-scale dynamo associated with planar averaged fields.
The minimum requirement for large-scale MRI dynamo growth has been shown to be
anistropic fluctuations and shear to form a nonzero EMF \citep{EB16}. 
The form of EMF in terms of a mean field theory, such as incoherent alpha-shear or helicity flux source, 
is yet to be investigated \citep{VB97,VC2001,EB2014}.
Unlike runs in elongated shearing boxes, along with the 
peak at $k=1$, there is a second peak around $k=4$--$5$.
Further investigation of the
secondary peak is beyond the scope of the present paper.

Lastly, we show the 
azimuthal ($B_y$) 
component of magnetic fields from Run~SSD, Run~A0 and Run~STD with a higher resolution of $512^3$ in \Fig{field}. 
For Run~SSD, most of the box is orange indicating weaker small-scale fields. 
The scattered appearance of yellow or blue regions (both indicating higher  magnitude fields) is due to the intermittency of the SSD. For  Run~A0 
and Run~STD, there are longer more coherent structures, particularly in the azimuthal direction, indicating the presence  of large-scale fields. 
Also there are stronger small-scale fields as well, indicating 
higher  contributions to Maxwell stresses  compared to Run~SSD. 
Note the characteristic vertical loopy structures (an MRI signature) in the $x$-$z$ plane in both Run~A0 and Run~STD. 
Thus Run~A0 and Run~STD compare well indicating that such simulations (with zero net flux) are independent of  the respective initial conditions.  

\section{Conclusions}

We have shown via direct numerical simulations that the MRI can sustain MHD turbulence even when seeded with an initially random  
small-scale magnetic and velocity fields supplied by an SSD.
In this case,  the energy is strongly dominated by fields at small scales, implying that 
substantial power at low wave modes is not necessary for MRI  growth

There is  however, still a  minimum strength and coherence required for this growth that is determined by a comparison between the turbulent diffusion time and  MRI growth time near
the wave number of maximum MRI growth.
When the latter time scale is shorter than the former,  diffusion wins and the MRI does not grow.
If the  turbulent velocity forcing scale 
(a proxy for coherence) of initial fields is decreased or the field magnitude is decreased (by supplying an early 
unsaturated SSD state as an initial condition) then the fastest growing linear MRI mode moves to
smaller scales where it has a harder time competing with diffusion.
In short, if the initial coherence scales or strengths are too small,
the MRI is quenched.

Generally, when the initial conditions are  supplied  by a saturated SSD, we find the conditions to be favorable for growth  
but when the initial field has a Gaussian random noise field the MRI fails to grow.  
The SSD is a natural initial condition as it is likely to be
common in sufficiently conducting plasmas as in stars or the galactic ISM
that feed accretion flows.
 
After the MRI takes over from the SSD supplied initial conditions, 
the saturated state of the magnetic and velocity fields in our
simulations is essentially indistinguishable from that 
of the more commonly studied case of initial vertical fields of  zero net flux. 
In particular the saturated amplitudes of the total magnetic and velocity fields (magnetic field being dominant), the accretion stresses, ratio of 
Maxwell to Reynolds stress, and the magnetic and kinetic power spectra are all 
very similar for the two aforementioned initial conditions.

Finally, we hypothesize that if only the  SSD produced tangled  magnetic field were supplied as an initial condition with the velocity removed,  
the turbulent decay rate would  be lowered because the pre-removal $U_{rms} \ge  V_A$.  In the absence of an initial velocity, 
the minimum decay time is an Alfv\'en crossing time, which is itself the MRI growth time at $k_{max}$.  
Thus the aforementioned minimum field strength and coherence requirements are, if anything, less stringent when the velocity is absent.  
Further exploration of this hypothesis, and the
understanding of the large-scale dynamo seen in these simulations
would be of interest to explore in future work.

\section*{Acknowledgments}
We thank Jim Stone and Greg Hammett for thoughtful questions and discussions and
Luca Comisso and Manasvi Lingam for some useful suggestions. 
PB and FE acknowledge grant support from DOE, DE-SC0012467.  
EB acknowledges support from  grants HST-AR-13916.002, and NSF AST1515648.
The computing resources were provided by Princeton Institute of Computational
Science (PICSciE).

\bibliographystyle{mnras}
\bibliography{ssdmri}

\label{lastpage}
\end{document}

%% file: macros.tex
\newcommand{\EQ}{\begin{equation}}
\newcommand{\EN}{\end{equation}}
\newcommand{\EQA}{\begin{eqnarray}}
\newcommand{\ENA}{\end{eqnarray}}

\newcommand{\Fig}[1]{Fig.~\ref{#1}}

\newcommand{\Tab}[1]{Table~\ref{#1}}

\newcommand{\bra}[1]{\langle #1\rangle}

{}
{}
{}

{}
{}
{}
{}
{}
{}
{}
{}
{}
{}
{}
{}
{}
{}
{}
{}
{}
{}
{}
{}
{}
{}

{}
{}
{}

{}

{}
{}
{}

\newcommand{\zzz}{\hat{\mbox{\boldmath $z$}} {}}

\newcommand{\UU}{\bm{U}}

\def\la{\mathrel{\mathchoice {\vcenter{\offinterlineskip\halign{\hfil
$\displaystyle##$\hfil\cr<\cr\sim\cr}}}
{\vcenter{\offinterlineskip\halign{\hfil$\textstyle##$\hfil\cr<\cr\sim\cr}}}
{\vcenter{\offinterlineskip\halign{\hfil$\scriptstyle##$\hfil\cr<\cr\sim\cr}}}
{\vcenter{\offinterlineskip\halign{\hfil$\scriptscriptstyle##$\hfil\cr<\cr\sim\cr}}}}}
\def\ga{\mathrel{\mathchoice {\vcenter{\offinterlineskip\halign{\hfil
$\displaystyle##$\hfil\cr>\cr\sim\cr}}}
{\vcenter{\offinterlineskip\halign{\hfil$\textstyle##$\hfil\cr>\cr\sim\cr}}}
{\vcenter{\offinterlineskip\halign{\hfil$\scriptstyle##$\hfil\cr>\cr\sim\cr}}}
{\vcenter{\offinterlineskip\halign{\hfil$\scriptscriptstyle##$\hfil\cr>\cr\sim\cr}}}}}
\def\Pm{\mbox{\rm Pr}_M}
\def\Rm{R_{\rm m}}

\def\EEK{{\cal E}_{\rm K}}
\def\EEM{{\cal E}_{\rm M}}

\def\kf{k_{\rm f}}

\def\kmax{k_{\rm max}}

\def\Brms{B_{\rm rms}}

\def\Urms{U_{\rm rms}}